\documentclass[aps,prl,twocolumn,showpacs,superscriptaddress,groupedaddress]{revtex4}
\usepackage{graphicx}  
\usepackage{dcolumn}   
\usepackage{bm}        
\usepackage{amsmath}   
\usepackage{amssymb}   
\usepackage{url}
\usepackage{datetime}  
\newcommand{\ttb}{\ensuremath{t\bar{t}}}

\newcommand{\invfb}{fb$^{-1}$}
\newcommand{\met}{\mbox{$\not\!\!E_{T}$}}

\newcommand{\nnt}{\ensuremath{\text{\emph{NN}$_\tau$}}}
\newcommand{\nnb}{\ensuremath{\text{\emph{NN}$\!_b$}}}
\newcommand{\nnd}{\ensuremath{\text{\emph{NN}$\!_{sb}$}}}
\newcommand{\geant}{{\sc geant}}
\newcommand{\tauh}{\ensuremath{\tau_h}}
\newcommand{\dmt}{\sigma_t}
\newcommand{\dmw}{\sigma_W}
\newcommand{\sttb}{\sigma_{\ttb}}
\newcommand{\BFtj}{\text{BR}_{\tauh+j}}
\begin{document}
\hspace{5.2in} \mbox{Fermilab-Pub-10-333-E}
\title{Measurement of $\boldsymbol{\ttb}$ production in the
  $\boldsymbol{\tau}$ + jets topology using
  $\boldsymbol{p\bar{p}}$ collisions at $\boldsymbol{\sqrt{s} = 1.96}$~TeV}
\affiliation{Universidad de Buenos Aires, Buenos Aires, Argentina}
\affiliation{LAFEX, Centro Brasileiro de Pesquisas F{\'\i}sicas, Rio de Janeiro, Brazil}
\affiliation{Universidade do Estado do Rio de Janeiro, Rio de Janeiro, Brazil}
\affiliation{Universidade Federal do ABC, Santo Andr\'e, Brazil}
\affiliation{Instituto de F\'{\i}sica Te\'orica, Universidade Estadual Paulista, S\~ao Paulo, Brazil}
\affiliation{Simon Fraser University, Vancouver, British Columbia, and York University, Toronto, Ontario, Canada}
\affiliation{University of Science and Technology of China, Hefei, People's Republic of China}
\affiliation{Universidad de los Andes, Bogot\'{a}, Colombia}
\affiliation{Charles University, Faculty of Mathematics and Physics, Center for Particle Physics, Prague, Czech Republic}
\affiliation{Czech Technical University in Prague, Prague, Czech Republic}
\affiliation{Center for Particle Physics, Institute of Physics, Academy of Sciences of the Czech Republic, Prague, Czech Republic}
\affiliation{Universidad San Francisco de Quito, Quito, Ecuador}
\affiliation{LPC, Universit\'e Blaise Pascal, CNRS/IN2P3, Clermont, France}
\affiliation{LPSC, Universit\'e Joseph Fourier Grenoble 1, CNRS/IN2P3, Institut National Polytechnique de Grenoble, Grenoble, France}
\affiliation{CPPM, Aix-Marseille Universit\'e, CNRS/IN2P3, Marseille, France}
\affiliation{LAL, Universit\'e Paris-Sud, CNRS/IN2P3, Orsay, France}
\affiliation{LPNHE, Universit\'es Paris VI and VII, CNRS/IN2P3, Paris, France}
\affiliation{CEA, Irfu, SPP, Saclay, France}
\affiliation{IPHC, Universit\'e de Strasbourg, CNRS/IN2P3, Strasbourg, France}
\affiliation{IPNL, Universit\'e Lyon 1, CNRS/IN2P3, Villeurbanne, France and Universit\'e de Lyon, Lyon, France}
\affiliation{III. Physikalisches Institut A, RWTH Aachen University, Aachen, Germany}
\affiliation{Physikalisches Institut, Universit{\"a}t Freiburg, Freiburg, Germany}
\affiliation{II. Physikalisches Institut, Georg-August-Universit{\"a}t G\"ottingen, G\"ottingen, Germany}
\affiliation{Institut f{\"u}r Physik, Universit{\"a}t Mainz, Mainz, Germany}
\affiliation{Ludwig-Maximilians-Universit{\"a}t M{\"u}nchen, M{\"u}nchen, Germany}
\affiliation{Fachbereich Physik, Bergische  Universit{\"a}t Wuppertal, Wuppertal, Germany}
\affiliation{Panjab University, Chandigarh, India}
\affiliation{Delhi University, Delhi, India}
\affiliation{Tata Institute of Fundamental Research, Mumbai, India}
\affiliation{University College Dublin, Dublin, Ireland}
\affiliation{Korea Detector Laboratory, Korea University, Seoul, Korea}
\affiliation{CINVESTAV, Mexico City, Mexico}
\affiliation{FOM-Institute NIKHEF and University of Amsterdam/NIKHEF, Amsterdam, The Netherlands}
\affiliation{Radboud University Nijmegen/NIKHEF, Nijmegen, The Netherlands}
\affiliation{Joint Institute for Nuclear Research, Dubna, Russia}
\affiliation{Institute for Theoretical and Experimental Physics, Moscow, Russia}
\affiliation{Moscow State University, Moscow, Russia}
\affiliation{Institute for High Energy Physics, Protvino, Russia}
\affiliation{Petersburg Nuclear Physics Institute, St. Petersburg, Russia}
\affiliation{Stockholm University, Stockholm and Uppsala University, Uppsala, Sweden }
\affiliation{Lancaster University, Lancaster LA1 4YB, United Kingdom}
\affiliation{Imperial College London, London SW7 2AZ, United Kingdom}
\affiliation{The University of Manchester, Manchester M13 9PL, United Kingdom}
\affiliation{University of Arizona, Tucson, Arizona 85721, USA}
\affiliation{University of California Riverside, Riverside, California 92521, USA}
\affiliation{Florida State University, Tallahassee, Florida 32306, USA}
\affiliation{Fermi National Accelerator Laboratory, Batavia, Illinois 60510, USA}
\affiliation{University of Illinois at Chicago, Chicago, Illinois 60607, USA}
\affiliation{Northern Illinois University, DeKalb, Illinois 60115, USA}
\affiliation{Northwestern University, Evanston, Illinois 60208, USA}
\affiliation{Indiana University, Bloomington, Indiana 47405, USA}
\affiliation{Purdue University Calumet, Hammond, Indiana 46323, USA}
\affiliation{University of Notre Dame, Notre Dame, Indiana 46556, USA}
\affiliation{Iowa State University, Ames, Iowa 50011, USA}
\affiliation{University of Kansas, Lawrence, Kansas 66045, USA}
\affiliation{Kansas State University, Manhattan, Kansas 66506, USA}
\affiliation{Louisiana Tech University, Ruston, Louisiana 71272, USA}
\affiliation{University of Maryland, College Park, Maryland 20742, USA}
\affiliation{Boston University, Boston, Massachusetts 02215, USA}
\affiliation{Northeastern University, Boston, Massachusetts 02115, USA}
\affiliation{University of Michigan, Ann Arbor, Michigan 48109, USA}
\affiliation{Michigan State University, East Lansing, Michigan 48824, USA}
\affiliation{University of Mississippi, University, Mississippi 38677, USA}
\affiliation{University of Nebraska, Lincoln, Nebraska 68588, USA}
\affiliation{Rutgers University, Piscataway, New Jersey 08855, USA}
\affiliation{Princeton University, Princeton, New Jersey 08544, USA}
\affiliation{State University of New York, Buffalo, New York 14260, USA}
\affiliation{Columbia University, New York, New York 10027, USA}
\affiliation{University of Rochester, Rochester, New York 14627, USA}
\affiliation{State University of New York, Stony Brook, New York 11794, USA}
\affiliation{Brookhaven National Laboratory, Upton, New York 11973, USA}
\affiliation{Langston University, Langston, Oklahoma 73050, USA}
\affiliation{University of Oklahoma, Norman, Oklahoma 73019, USA}
\affiliation{Oklahoma State University, Stillwater, Oklahoma 74078, USA}
\affiliation{Brown University, Providence, Rhode Island 02912, USA}
\affiliation{University of Texas, Arlington, Texas 76019, USA}
\affiliation{Southern Methodist University, Dallas, Texas 75275, USA}
\affiliation{Rice University, Houston, Texas 77005, USA}
\affiliation{University of Virginia, Charlottesville, Virginia 22901, USA}
\affiliation{University of Washington, Seattle, Washington 98195, USA}
\author{V.M.~Abazov} \affiliation{Joint Institute for Nuclear Research, Dubna, Russia}
\author{B.~Abbott} \affiliation{University of Oklahoma, Norman, Oklahoma 73019, USA}
\author{M.~Abolins} \affiliation{Michigan State University, East Lansing, Michigan 48824, USA}
\author{B.S.~Acharya} \affiliation{Tata Institute of Fundamental Research, Mumbai, India}
\author{M.~Adams} \affiliation{University of Illinois at Chicago, Chicago, Illinois 60607, USA}
\author{T.~Adams} \affiliation{Florida State University, Tallahassee, Florida 32306, USA}
\author{G.D.~Alexeev} \affiliation{Joint Institute for Nuclear Research, Dubna, Russia}
\author{G.~Alkhazov} \affiliation{Petersburg Nuclear Physics Institute, St. Petersburg, Russia}
\author{A.~Alton$^{a}$} \affiliation{University of Michigan, Ann Arbor, Michigan 48109, USA}
\author{G.~Alverson} \affiliation{Northeastern University, Boston, Massachusetts 02115, USA}
% ...
%
%
\collaboration{The D0 Collaboration\footnote{with visitors from
$^{a}$Augustana College, Sioux Falls, SD, USA,
$^{b}$The University of Liverpool, Liverpool, UK,
$^{c}$SLAC, Menlo Park, CA, USA,
$^{d}$ICREA/IFAE, Barcelona, Spain,
$^{e}$Centro de Investigacion en Computacion - IPN, Mexico City, Mexico,
$^{f}$ECFM, Universidad Autonoma de Sinaloa, Culiac\'an, Mexico,
and 
$^{g}$Universit{\"a}t Bern, Bern, Switzerland.%
}} \noaffiliation
\vskip 0.25cm

\date{August 24, 2010}

\begin{abstract}
We present a measurement of the \ttb\ production cross section
multiplied by the branching ratio to tau lepton decaying
semihadronically ($\tauh$) plus jets,
$\sigma(p\bar{p}\to \ttb+X)
\cdot \text{BR}(\ttb\to\tauh+\text{jets})$,
at a center of mass energy $\sqrt{s}=1.96$~TeV using
$1$~fb$^{-1}$ of integrated luminosity collected with the D0
detector. Assuming a top quark mass of 170~GeV,
we measure
$\sttb\cdot\BFtj=0.60^{+0.23}_{-0.22}\;(\text{stat})\;^{+0.15}_{-0.14}\;
  (\text{syst})\pm 0.04\;(\text{lumi})\ \text{pb}$. In addition,
we extract the \ttb\ production cross section using
the $\ttb\to\tauh+\text{jets}$ topology, with the result
$\sttb =
  6.9\;_{-1.2}^{+1.2}\;({\textrm{stat}})\;_{-0.7}^{+0.8}\;({\text{syst}})
  \pm 0.4\;({\text{lumi}})\ \text{pb}$. These findings are in
good agreement with standard model
predictions and measurements performed using other top quark decay channels.
\end{abstract}

\maketitle 

The decay $t\to Wb\to\tau\nu_{\tau}b$ provides a unique laboratory
to investigate the properties of the third generation fermions ---
the top ($t$) and bottom ($b$) quarks, the tau lepton ($\tau$),
and the tau neutrino ($\nu_{\tau}$) --- in a single process. In the
standard model (SM), the $t$ quark branching ratio (BR) to a $W$
boson and a $b$ quark is $\approx100\%$, and the final state is determined
by the SM BR of the $W$ boson. Since the {$t$ }is the heaviest quark
and the $\tau$ the heaviest lepton, any non-SM mass- or flavor-dependent
couplings could change the $t$ quark decay rate into final states
with $\tau$ leptons. 
Therefore, any deviation in the
BR of $t\to\tau\nu_{\tau}b$ from that predicted by the SM can
be an indication of non-SM physics. For example, in
the Type 2 two-Higgs doublet model~\cite{THDM}, such as required
by the minimal supersymmetric standard model~\cite{MSSM}, the $t$
quark can have a significant BR to a charged Higgs boson ($H^\pm$)
and a $b$ quark
if $m_{H^\pm}<m_t-m_b$.
For large values of $\tan\beta$, the ratio of the vacuum expectation
values of the two-Higgs doublets, the charged Higgs boson preferentially
decays to $\tau\nu_{\tau}$, thereby increasing the
BR of $t\to\tau\nu_{\tau}b$ relative to the SM expectation and
leading to a larger measured
$\sigma(p\bar{p}\to\ttb+X)\cdot\text{BR}(\ttb\to\tau+\text{jets})$
compared to the value expected
from SM assumptions for the BRs and the production cross
section~\cite{ttbtheory1, ttbtheory2, ttbtheory3}.  Other
possible non-SM processes that can enhance the $t$ quark to $\tau$ lepton
BR are $R$-parity violating decays of the $t$ quark in supersymmetric
models~\cite{Rsusy}
and new $Z'$ bosons with nonuniversal couplings~\cite{Zprime}. 

In this article, we present the first measurement of $\ttb$
production in the $\tau+\text{jets}$ final state using a data sample
corresponding to an integrated luminosity of
$1$~\invfb\ collected with the D0 detector~\cite{d0det}
at the Fermilab Tevatron $p\bar{p}$ Collider
operating at a center of mass energy $\sqrt{s}=1.96$~TeV. This
measurement uses semihadronic $\tau$ lepton decays, with
$\text{BR}\approx 65\%$, as secondary electrons
and muons from $\tau$ lepton decays are difficult to distinguish from
primary electrons and muons resulting from $W$ decays. Previous
measurements of \ttb\ production using $\tau$ leptons in the
final state have been performed by the D0~\cite{d0taulep} and
CDF~\cite{cdftaulep} collaborations in the $\tauh+\ell$ channel, where
$\tauh$ represents semihadronic $\tau$ lepton decay modes and
$\ell$ represents either an electron or a muon.

We apply the following preselection requirements:
events must satisfy a multijet trigger
requiring at least four jets; this is the same trigger used in the
\ttb\ cross section measurement in the all-hadronic decay
mode~\cite{d0alljets}. 
Reconstructed events are required to have
missing transverse energy $\met\ge 15$~GeV and
$\met\ \text{significance}>3$,
where the $\met\ \text{significance}$ is a measure
of the likelihood that the
$\met$ arises from physical sources
rather than fluctuations in the measurement of the energies of
the physics objects (jets, muons, 
electrons and unclustered energy)~\cite{metl}. 
Each event must also have at least four
reconstructed jets with pseudorapidity $|\eta|<2.5$
and transverse momentum
$p_T>15$~GeV using an iterative jet
cone algorithm~\cite{d0jets} with a cone size 
$\Delta\mathcal{R}=\sqrt{(\Delta\eta)^2+(\Delta\phi)^2}=0.5$~\cite{prapid}.
The jet energies are corrected for the energy response of
the calorimeter, the cone size, multiple $p\bar{p}$ interactions, event
pile-up, and calorimeter noise~\cite{jes}.
At least one jet is required to have $p_T>35$~GeV,
and at least two jets are required to have $p_T>25$~GeV.
Each event is also required
to have at
least one $\tauh$ candidate with $p_T>10$~GeV, $|\eta|<2.5$,
and tau neural network output, $\nnt>0.3$~\cite{tauNN}. Finally,
to ensure this analysis is statistically independent of other
D0 \ttb\ cross section measurements so that it can be included in
a combined cross section measurement,
events satisfying the requirements of the $\ttb\to e(\mu)+\text{jets}$
channel~\cite{isol}, which include an isolated electron (muon) with
$p_T>20$~GeV, are rejected, as are events satisfying the requirements of
the \ttb\ cross section measurement
in the all-hadronic channel~\cite{d0alljets}.

A semihadronic $\tau$ lepton candidate is a calorimeter cluster of cone
size $\Delta\mathcal{R}=0.5$ that includes any
subclusters that might be present with $E>800$~MeV constructed from
cells in the electromagnetic (EM) section of the calorimeter and the
associated tracks with $p_T>1.5$~GeV in a cone
$\Delta\mathcal{R}= 0.3$ contained within the calorimeter cluster.
These $\tau$ candidates are classified according to one of
three types based on the number of tracks and activity
in the EM calorimeter, motivated by 
the semihadronic $\tau$ lepton decays:
(1) $\tau^\pm\to\pi^\pm\nu_\tau$, (2) $\tau^\pm\to
\pi^\pm\pi^0\nu_\tau$, (3) $\tau^\pm\to\pi^\pm\pi^\pm\pi^\mp(\pi^0)\nu_\tau$.
We define the three tau-types as follows: a single track with
no EM subclusters (tau-type 1);  a single track and $\ge 1$~EM
subclusters (tau-type 2); and at least two tracks and $\ge 0$~EM 
subclusters (tau-type 3).

To further
reduce the number of quark and gluon jets reconstructed as $\tau$ leptons,
we train separate neural networks for each
$\tauh$ lepton decay type to improve the discrimination
of $\tau$ lepton candidates from the jet background.
The input variables
to \nnt\ are chosen to be minimally dependent on the $\tau$ lepton energy
and to exploit the low track multiplicity and
the narrow width of the calorimeter cluster produced by
$\tau$ leptons decaying semihadronically, the low mass of the $\tau$
lepton, and
the differences in longitudinal and transverse shower shapes between
$\tau$ leptons and jets~\cite{tauNN}.
A total of 12 $\nnt$ input variable are used to characterize
the presence and properties of $\tauh$ leptons, with seven of these
variables in common for all three tau-types. The 12 variables
are classified as follows: isolation variables, shower shape variables,
and correlation variables between the calorimeter cluster and the
associated charged particle tracks.
Each $\nnt$
is trained on $Z\to \tau^+\tau^-$ Monte Carlo (MC) events for signal
and jets from data, where a jet and a nonisolated muon are back-to-back
in $\phi$, for background. 
These are the same training samples used in Ref~\cite{ztau}.

To measure the number of
$\ttb\to\tauh+\text{jets}$ signal events in data, the
physics and instrumental backgrounds must be determined.
The main physics backgrounds are
$W+\text{jets}$ events, where the $W$ boson decays to a $\tau$ lepton,
and to a smaller extent $Z+\text{jets}$ events, where the $Z$ boson
decays to a pair of $\tau$ leptons with one misidentified as
a jet and the \met\ is due to the neutrinos from the
decays of the $\tau$ leptons.
The
main instrumental background is multijet production
where a jet is misidentified as a $\tau$ lepton and
the energy is mismeasured leading to a net \met.

The preselection efficiencies and SM BRs for \ttb\ to final states with
leptons~\cite{pdg} are given in Table~\ref{tab:presel}. These, as well as 
the final efficiencies, are calculated using a MC simulation of
the experiment. The \ttb\ signal with leptons in the final state
and $W(Z)+\text{jets}$ background are simulated using
the \textsc{alpgen~1.2}~\cite{alpgen} matrix element generator
assuming a $t$ quark mass 
of 170~GeV and using the CTEQ6L1~\cite{cteq}
parton distribution function set. These events are then processed 
through \textsc{pythia~6.2}~\cite{pythia} to simulate parton showering,
fragmentation, hadronization, and decays of short lived particles,
except for $b$ hadrons and $\tau$ leptons. 
\textsc{evtgen} \cite{evtgen} is used to model the decays of $b$ hadrons, while 
$\tau$ leptons are decayed using \textsc{tauola}~\cite{tauola}.
To avoid double counting final states generated by the leading-order
parton-level calculation of \textsc{alpgen} and the parton-level
shower evolution of \textsc{pythia}, a matching algorithm is
used~\cite{dcount}.
The generated 
events are then processed through the \geant-based~\cite{geant} simulation 
of the D0 detector providing tracking hits, calorimeter cell 
energies and muon hit information. The 
same reconstruction algorithm is applied to data and simulated events.
\begin{table}
\caption{A summary of the SM BRs of the various \ttb\ 
  subprocesses
  and the preselection efficiencies, where
  the uncertainties are derived from MC statistics.
  The
  leptonic $\tau$ lepton decays are included in the $e$ and $\mu$ channels, and
  $l^\pm$ represents an $e$, $\mu$ or $\tau$ lepton. }
\begin{ruledtabular}
\begin{tabular}{lD{.}{.}{-1}D{,}{\pm}{-1}} 
&\multicolumn{1}{c}{BR (\%)}
&\multicolumn{1}{c}{$\epsilon_\text{preselection}$ (\%)}
\\
\hline
$\ttb\to\tauh+\text{jets}$&
9.75&
40.5 , 0.2
\\
$\ttb\to e+\text{jets}$&
17.7&
17.0 , 0.2
\\
$\ttb\to \mu+\text{jets}$&
17.6&
11.1 , 0.1
\\
$\ttb\to l^+l^-+\text{jets}$&
11.1&
4.04 , 0.03
\\
\end{tabular}
\end{ruledtabular}
\label{tab:presel}
\end{table}

The preselected data sample is used
to extract the signal and to study the multijet background after
additional selection criteria are applied. To extract
the signal sample, we require $\nnt>0.95$.
The selected events are then separated on the basis of
tau-type according to the $\tau$ lepton candidate with the highest value
of \nnt.
This is done primarily to separate tau-type 3 events
from the tau-type 1 and 2 events, since the former has a much higher
misidentification rate and thus result in larger uncertainties on the \ttb\ 
cross section.
In addition, we require
that each event have at least one identified $b$ jet using the
$b$-tag neural network (\nnb) with the requirement $\nnb>0.775$. The \nnb\ uses
nine input variables that characterize the presence and properties of
secondary vertices and track impact parameters within the 
jet~\cite{btag}.
The efficiencies of these selections are shown in
Table \ref{tight}.

\begin{table*}
\caption{The efficiencies for the tight $\tau$ lepton
  candidate ($\nnt > 0.95$) and $b$-tagging
  selections for tau-type 1 and 2, and tau-type 3 channels.
  The uncertainties are based on MC statistics.}
\renewcommand{\arraystretch}{1.25}
\begin{ruledtabular}
\begin{tabular}{lcD{,}{\pm}{-1}ccD{,}{\pm}{-1}c}
&\multicolumn{1}{c}{Tau-types 1 and 2}
&\multicolumn{1}{c}{Tau-types 1 and 2}
&\multicolumn{1}{c}{Tau-types 1 and 2}
&\multicolumn{1}{c}{Tau-type 3}
&\multicolumn{1}{c}{Tau-type 3}
&\multicolumn{1}{c}{Tau-type 3}\\
&\multicolumn{1}{c}{Trigger (\%)}
&\multicolumn{1}{c}{$\nnt>0.95$ (\%)}
&\multicolumn{1}{c}{$b$-tag (\%)}
&\multicolumn{1}{c}{Trigger (\%)}
&\multicolumn{1}{c}{$\nnt>0.95$ (\%)}
&\multicolumn{1}{c}{$b$-tag (\%)}
\\
\hline
$\ttb\rightarrow\tauh+\text{jets}$&
$74.8^{+1.7}_{-0.1}$&
23.7 , 0.3&
$60.1_{-2.7}^{+2.8}$&
$73.6^{+1.6}_{-0.1}$& 
19.4 , 0.2&
$59.9_{-2.7}^{+2.8}$
\\
$\ttb\rightarrow e+\text{jets}$&
$69.9^{+1.5}_{-0.1}$&
33.1 , 0.4&
$58.7_{-2.7}^{+2.8}$&
$66.0^{+1.5}_{-0.1}$& 
8.1 , 0.2&
$58.9_{-2.7}^{+2.8}$
\\
$\ttb\rightarrow \mu+\text{jets}$&
$65.9^{+1.5}_{-0.1}$&
3.8 , 0.1&
$60.3_{-2.7}^{+2.8}$&
$66.1^{+1.4}_{-0.1}$& 
7.7 , 0.2&
$59.0_{-2.7}^{+2.8}$
\\
$\ttb\rightarrow l^+l^-+\text{jets}$&
$50.5^{+1.1}_{-0.2}$&
43.7 , 0.4&
$60.2_{-2.7}^{+2.8}$&
$50.7^{+1.1}_{-0.2}$& 
20.6 , 0.3&
$61.4_{-2.8}^{+2.9}$
\\
\end{tabular}
\end{ruledtabular}
\label{tight} 
\end{table*}
The expected fraction of \ttb\ events in the signal sample
is $\approx 15\%$ for tau-type 1 and 2, and $\approx 3\%$
for tau-type~3 assuming
$\sttb=6.9$~pb as measured in this
analysis. In addition,
the signal sample contains $W(Z)+\text{jets}$
and multijet background events that must be subtracted.
The $W(Z)+\text{jets}$ contamination is determined using MC events, while
the multijet background is determined from data. We start with
the preselected sample and apply a loose $\tau$ lepton veto,
$\nnt<0.9$. Using MC events, we expect that the resulting sample
contains $< 2\%$
$\ttb\to\tauh+\text{jets}$ events and $<3\%$ $W(Z)+\text{jets}$
events, and therefore provides a good representation of the
multijet background. To further improve the modeling,
the $W(Z)+\text{jets}$ expectation is subtracted from
the multijet background data sample.

The numbers of signal and background events
are extracted from the final selected sample
using a neural network (\nnd) event discriminant with the following
input variables:
(1) the scalar sum of the $p_T$ of all jets and the $\tau$ lepton
candidate
in the event; (2) the aplanarity~\cite{aplan};
(3) the \met\ significance;
(4) the invariant mass of all jets and the $\tau$ lepton candidate
in the event; and (5) a $\chi^2$ representing
how well the 2 and 3 jet invariant masses agree with
values expected for hadronic $t$ quark decays,
$\chi^2=(M_{\text{3jet}}-m_t)/\dmt^2
+(M_{\text{2jet}}-m_W)^2/\dmw^2$,
with 
$M_{\text{2jet}}$ ($M_{\text{3jet}}$) being the 2 (3) jet invariant mass,
$m_t=170$~GeV, $\dmt=45$~GeV and $m_W=80$~GeV, $\dmw=10$~GeV
are the mass and its resolution in the all-hadronic final state
for the $t$ quark and $W$ boson,
respectively.
The jet combination minimizing the $\chi^2$ is used.
The \nnd\ is trained using
a generated $\ttb\to\tauh+\text{jets}$ MC sample
for signal and half the multijet data sample for background.

\begin{figure}
\centerline{
\includegraphics[width=0.49\textwidth]{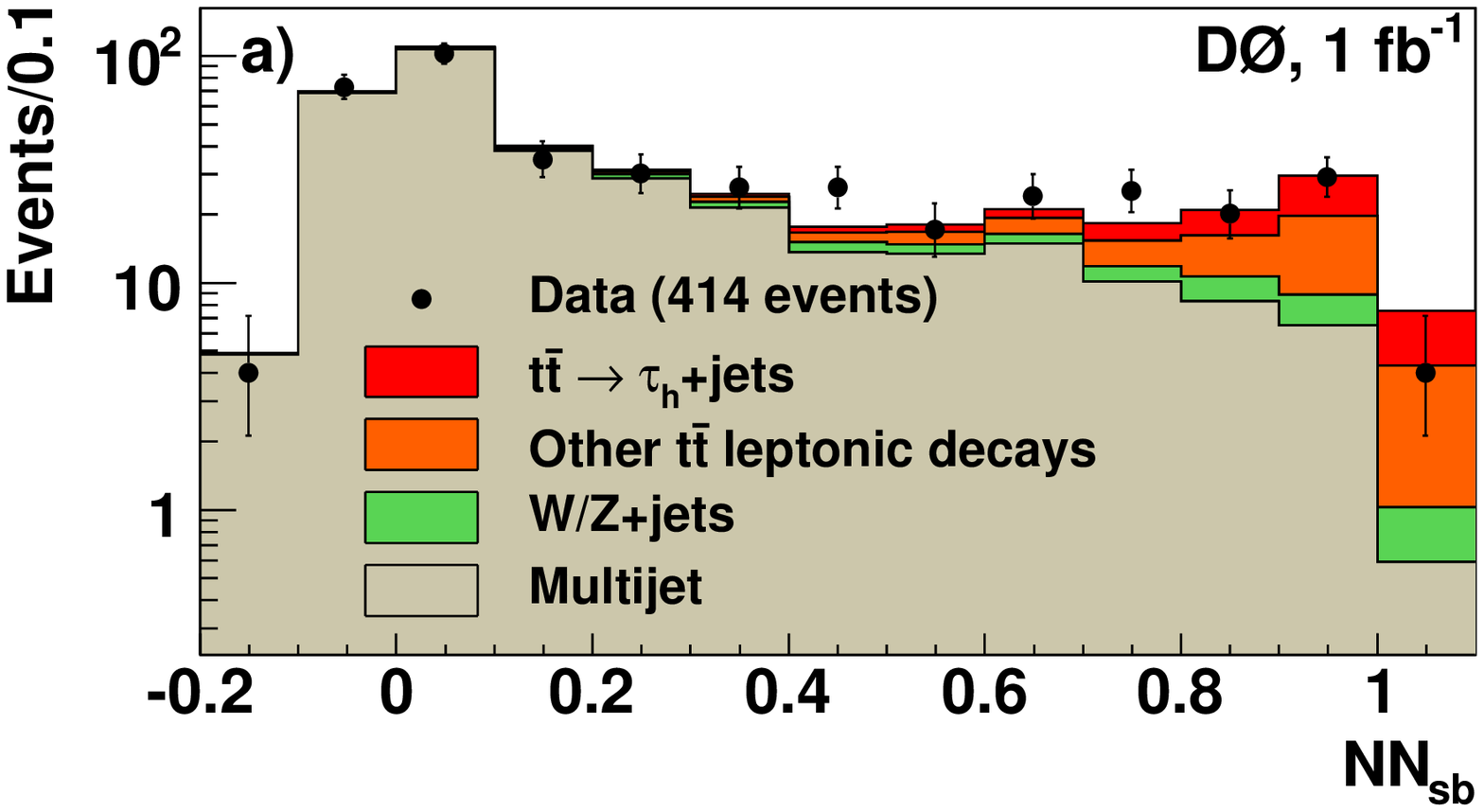}}
\centerline{
\includegraphics[width=0.49\textwidth]{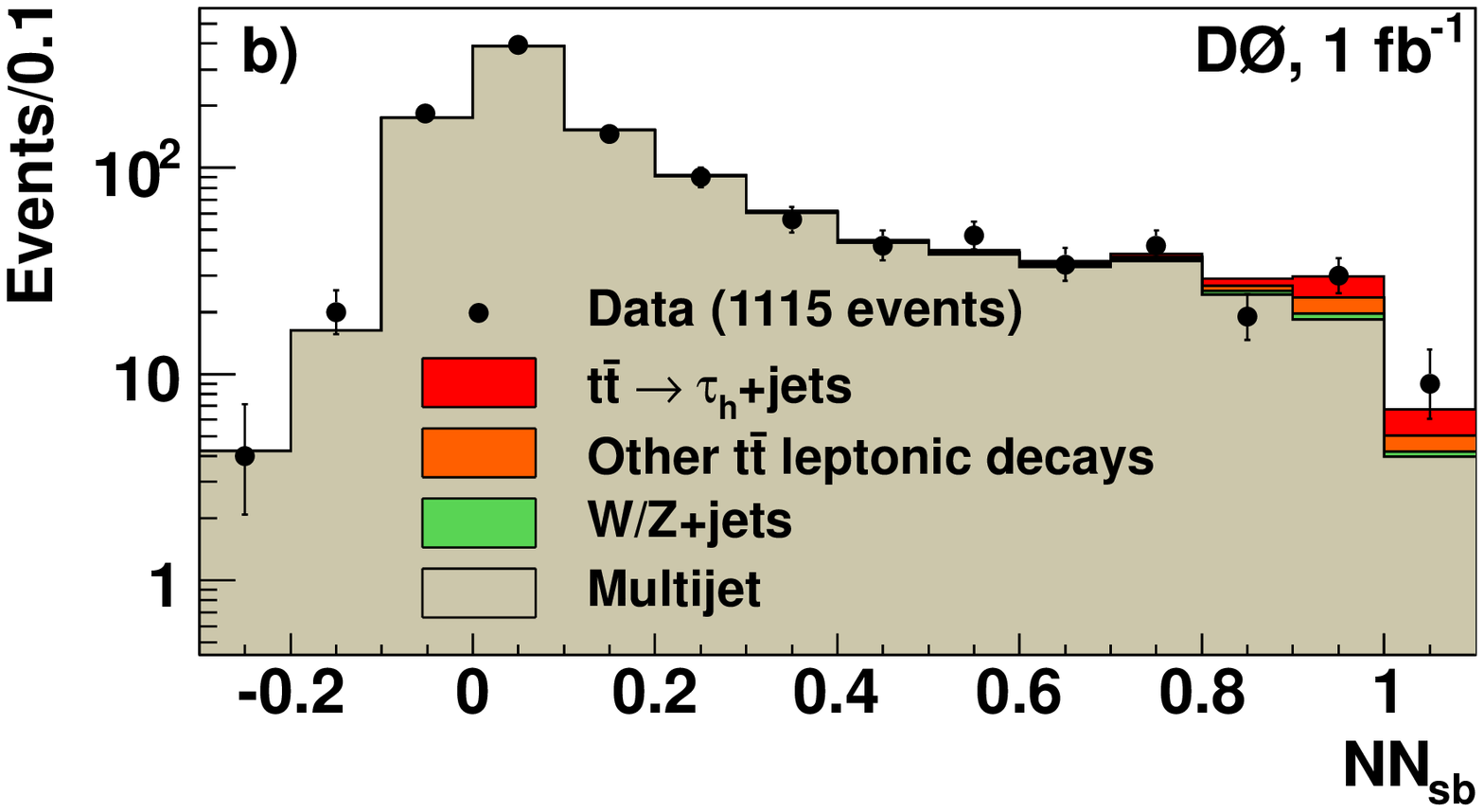}}
\caption{The output of \nnd\ for a) the tau-type 1 and 2
  channel, b) the tau-type 3 channel. The $\chi^2$ per degree
  of freedom between data and templates is 0.6 for a) and 0.5 for
  b).}
\label{fig:type12}
\end{figure}

We apply the trained \nnd\ to the signal data sample, the
remaining half of the multijet sample, a \ttb\ MC sample with
leptons in the final state that
is independent of the \nnd\ training sample,
and a $W(Z)+\text{jets}$ MC sample. The application of
\nnd\ on the multijet and MC samples is used to generate
templates,  as shown in
Fig.~\ref{fig:type12}, that are used
to determine the fraction of
\ttb\ and multijet events using a negative log-likelihood fit.
The normalization of the $W(Z)+\text{jets}$ MC sample is derived by
scaling the $W(Z)$ transverse (dilepton) mass distribution to data.
The normalization for
$\ttb\to e(\mu)+\text{jets}$ is fixed to the theoretical
cross section~\cite{ttbtheory3} and BRs.

The number of $\ttb\to\tauh+\text{jets}$ events extracted
from the fit to data are $25.1^{+11.2}_{-10.5}$~(stat) and $18.0^{+11.3}_{-10.5}$~(stat)
for channels with
tau-types 1 and 2 together, and with tau-type 3, respectively. The
fitted numbers of the multijet background events are
$336.4^{+11.2}_{-10.5}$~(stat) and $1083.2^{+11.3}_{-10.3}$~(stat),
for the two channels,
respectively. The numbers of \ttb\ events are comparable to the
expected values given in Table~\ref{tab:expct}.
\begin{table}
\caption{Expected event yields in the two analysis channels
  assuming the measured \ttb\ production cross section of 6.9~pb.
  The uncertainties are derived from MC statistics.}
\begin{ruledtabular}
\begin{tabular}{lD{,}{\pm}{4}D{,}{\pm}{4}}
&\multicolumn{1}{c}{Tau-type 1 and 2}
&\multicolumn{1}{c}{Tau-type 3}\\
\hline 
$\ttb\rightarrow\tauh+\text{jets}$&
27.6 , 0.4&
22.1 , 0.3\\
$\ttb\rightarrow e+\text{jets}$&
26.3 , 0.4&
5.9 , 0.2\\
$\ttb\rightarrow\mu+\text{jets}$&
2.0 , 0.1&
3.7 , 0.1\\
$\ttb\rightarrow l^+l^-+\text{jets}$&
4.1 , 0.1&
2.0 , 0.1\\\hline
Total 
$\ttb\to\text{leptons}$&
61.3, 0.6&
34.4, 0.4
\\\hline
$W$+jets&
13.5 , 0.3&
5.9 , 0.2\\
$Z$+jets&
3.4 , 0.4&
1.9 , 0.1\\
\end{tabular}
\end{ruledtabular}
\label{tab:expct} 
\end{table}

To minimize the statistical uncertainty of the measurement of
$\sigma(p\bar{p}\to\ttb+X)\cdot\text{BR}(\ttb\to\tauh+\text{jets})$,
which we denote as $\sttb\cdot\BFtj$, we fit the entire
\nnd\ output distribution rather than counting events
above a given value. The value of $\sttb\cdot\BFtj$ and the
fraction of multijet background in the sample are
obtained from a negative log-likelihood fit to the \nnd\ distributions
for tau-types 1 and 2 and tau-type 3, independently:
\begin{equation}
L(\sttb, \tilde{N}_{i}, N^{\text{obs}}_{i}) =
-\log\left(\prod_{i} \frac{\tilde{N}^{N^{\text{obs}}_{i}}_{i}}
  {N^{\text{obs}}_{i}!}  e^{-\tilde{N}_{i}}\right),\label{eq:nll}
\end{equation}
where \(\tilde{N}_{i} = \sttb \times \sum_{j} 
\epsilon_{\ttb(j)}^i \times \text{BR}_{\ttb(j)}
\times \mathcal{L}  + N_{\text{bkg}, i}\) is the expected number of
events in the $i^{th}$ bin of the \nnd\ histogram for a given
$\sttb$, with integrated luminosity $\mathcal{L}$, number of
background events $ N_{\text{bkg}, i}$, and 
the efficiency (BR) for the $j^{th}$ \ttb\
leptonic channel $\epsilon_{\ttb(j)}$ ($\text{BR}_{\ttb(j)}$), and
\(N^{\text{obs}}_{i}\) is the observed number of events in the $i^{th}$ bin.

The measured value of $\sttb\cdot\BFtj$ is
\begin{displaymath}
  0.60^{+0.23}_{-0.22}\;(\text{stat})\;^{+0.15}_{-0.14}\;
  (\text{syst})\pm 0.04\;(\text{lumi})\ \text{pb},
\end{displaymath}
where we combine the
tau-type 1 and 2 measurement with the tau-type 3 measurement. Using the
theoretical cross section $\sttb=8.06^{+0.52}_{-0.73}$~pb for
$m_t=170$~GeV from Ref.~\cite{ttbtheory3}, we measure
$\BFtj=0.074^{+0.029}_{-0.027}$ which
is consistent with the SM value given in Table~\ref{tab:presel}.

Table~\ref{cap:Syst} summarizes the systematic uncertainties on
$\sttb\cdot\BFtj$.
These are calculated by
varying the source by plus and minus 
one standard deviation,
and propagating the uncertainty to the final $\sttb\cdot\BFtj$.
The jet energy
corrections account for the effect of the jet energy scale and resolution.
Jet identification takes account of the difference in the jet
finding efficiency in data and MC. The $b$-tagging entry accounts
for the systematic uncertainties on its efficiency.
The $\tau$ lepton identification uncertainty is derived by fluctuating
the value of each input variable within its statistical
uncertainty and observing its effect on the \nnt\ output.
The trigger category accounts for the uncertainty in the multijet trigger
turn-on and also takes into account the possibility that a multijet event with
a $\tau$ lepton can have a different trigger turn-on.
Multijet modeling accounts for the uncertainty of
the multijet sample to model the $\ttb\to\tauh+\text{jets}$
background and its limited statistics. The category MC
modeling accounts for the $W+\text{jets}$ modeling, the
uncertainty in the scale factor both for light flavor
jets and heavy flavor jets, and the parton distribution function uncertainty.
The \ttb\ cross section systematic uncertainty represents
the effect of the normalization of the non-tau lepton \ttb\ background, which
is normalized to the theoretical value of the cross section.
In addition to the sources listed
in Table~\ref{cap:Syst}, there is a $\pm 6.1\%$ uncertainty in the
luminosity measurement~\cite{d0lumi}.
\begin{table*}
\caption{Systematic uncertainties on $\sttb\cdot\BFtj$ (in pb)
  as measured for the $\ttb\to\tauh+\text{jets}$ channel.}
\begin{ruledtabular}
\begin{tabular}{lcccccc}
Source&
\multicolumn{2}{c}{$\tauh$+jets (types 1 and 2)} &
\multicolumn{2}{c}{$\tauh$+jets (type 3)} &
\multicolumn{2}{c}{Combined} \\
\hline
Jet energy corrections&
$-0.078$& $+0.081$ &
$-0.047$& $+0.047$ &
$-0.068$& $+0.069$ \\
Jet identification &
$-0.019$& $+0.019$ &
$-0.012$& $+0.012$ &
$-0.016$& $+0.016$ \\
$b$ tagging &
$-0.074$& $+0.084$ &
$-0.035$& $+0.041$ &
$-0.060$& $+0.068$ \\
Tau identification&
$-0.035$& $+0.035$ &
$-0.020$& $+0.021$ &
$-0.029$& $+0.029$ \\
Trigger &
$-0.002$& $+0.053$ &
$-0.000$& $+0.027$ &
$-0.002$& $+0.043$ \\
Multijet modeling &
$-0.090$& $+0.090$ &
$-0.169$& $+0.169$ &
$-0.083$& $+0.083$ \\
MC modeling&
$-0.028$& $+0.028$ &
$-0.012$& $+0.013$ &
$-0.023$& $+0.022$ \\
\ttb\ cross section&
$-0.064$& $+0.068$ &
$-0.029$& $+0.030$ &
$-0.052$& $+0.055$ \\
\hline
Total systematic uncertainty&
$-0.16$ & $+0.15$ & $-0.18$ & $+0.15$& $-0.14$ & $+0.15$
\end{tabular}
\end{ruledtabular}
\label{cap:Syst} 
\end{table*}

In addition, we present the combined measurement of the production
cross section for \ttb\ using all measured \ttb\ channels with leptons
in the final state listed in Table~\ref{tab:expct}
that satisfy the selection criteria described above.
We repeat the
negative log-likelihood fit for the number of \ttb\ signal and multijet
background events fixing the \ttb\ BRs to their SM values, but this time
fit for all \ttb\ channels
arriving at $60.5 \pm 11.8$~(stat) events and $24.0 \pm 11.4$~(stat)
events for channels with tau-types 1 and 2 and with tau-type 3
characteristics, respectively. The fitted multijet backgrounds in this
case are $336.7 \pm 11.8$~(stat) events and $1083.2\pm 11.4$~(stat)
events, for the two channels,
respectively. The production cross section is calculated using the negative
log-likelihood defined in Eq.~\ref{eq:nll} for tau-types 1 and 2 and
tau-type 3 separately. The two cross sections are then combined to give
\begin{displaymath}
  \sttb = 
  6.9\;_{-1.2}^{+1.2}\;({\text{stat}})\;_{-0.7}^{+0.8}\;
  ({\text{syst}})\pm 0.4\;({\text{lumi}})\ \text{pb}.
\end{displaymath}
To estimate the dependence on $m_t$,
we reevaluate the efficiencies and templates using $m_t=175$~GeV
and find
\begin{displaymath}
  \sttb=6.3\;_{-1.1}^{+1.2}\;({\text{stat}})\;\pm _{-0.7}^{+0.7}\;
  ({\text{syst}})\pm 0.4\;({\text{lumi}})\ \text{pb}.
\end{displaymath}

In summary,
we have performed a measurement of $\sttb\cdot\BFtj=
0.60^{+0.28}_{-0.26}\ \text{pb}$ and,
using the theoretical \ttb\ production
cross section, extracted $\BFtj=0.074^{+0.029}_{-0.027}$, which agrees with
the SM expectation.
In addition, we have performed a measurement of the $p\bar{p}\to\ttb+X$
production cross section, $\sttb = 
6.9\;_{-1.4}^{+1.5}\ \text{pb}$,
using the $\ttb\to\tauh+\text{jets}$
topology. The measurement is in agreement with the
SM~\cite{ttbtheory1, ttbtheory2, ttbtheory3} and
previous experimental measurements using other \ttb\ 
channels~\cite{pdg} at the Tevatron.

%
We thank the staffs at Fermilab and collaborating institutions,
and acknowledge support from the
DOE and NSF (USA);
CEA and CNRS/IN2P3 (France);
FASI, Rosatom and RFBR (Russia);
CNPq, FAPERJ, FAPESP and FUNDUNESP (Brazil);
DAE and DST (India);
Colciencias (Colombia);
CONACyT (Mexico);
KRF and KOSEF (Korea);
CONICET and UBACyT (Argentina);
FOM (The Netherlands);
STFC and the Royal Society (United Kingdom);
MSMT and GACR (Czech Republic);
CRC Program and NSERC (Canada);
BMBF and DFG (Germany);
SFI (Ireland);
The Swedish Research Council (Sweden);
and
CAS and CNSF (China).

\begin{thebibliography}{99}
\bibitem{prapid} The D0 coordinate system has the positive 
$z$-axis along the proton beamline, and $z = 0$ at the center 
of the detector. The polar and azimuthal angles are denoted as 
$\theta$ and $\phi$, respectively. The pseudorapidity is defined 
as $\eta = - \ln(\tan \frac{\theta}{2})$.
\bibitem{aplan} The aplanarity is $3/2\lambda_3$, with $\lambda_3$
  being the smallest eigenvalue of the momentum tensor
  $M^{\alpha\beta}=\sum_ip_i^\alpha p_i^\beta/\sum_i|\vec{p}_i|^2$, where
  $i$ runs over the number of jets and the $\tau$ lepton candidate,
  and $\alpha, \beta=1,2,3$
  specifies the three spatial components of the momentum.
\end{thebibliography}
\end{document}